\documentstyle[12pt,fullpage,epsf]{article}
\catcode `@=11 \@addtoreset{equation}{section} \catcode `@=12

\def\gsim{\:\raisebox{-0.5ex}{$\stackrel{\textstyle>}{\sim}$}\:}

\def\be{\begin{equation}}
\def\ee{\end{equation}}
\def\beq{\begin{eqnarray}}
\def\eeq{\end{eqnarray}}
\def\T{{\cal T}}

\begin{document}
\setlength{\baselineskip}{2\baselineskip}
\hfill TIFR/TH/95-58
\bigskip

\begin{center}
{\large\bf Conservation Laws and Integrability of a One-dimensional
Model of Diffusing Dimers}
\bigskip

{\large Gautam I. Menon, Mustansir Barma and Deepak Dhar}

\medskip

{\large\em Theoretical Physics Group \\ Tata Institute of Fundamental
Research \\ Homi Bhabha Road, Mumbai 400 005, INDIA}

\end{center}

\bigskip 
\bigskip 
\bigskip 
\begin{center} {\bf Abstract} 
\end{center} 

We study a model of assisted diffusion of hard-core particles on a line.
Our model is a special case of a multispecies exclusion process, but the
long-time decay of correlation functions can be qualitatively different
from that of the simple exclusion process, depending on initial
conditions.  This behaviour is a consequence of the existence of an
infinity of conserved quantities. The configuration space breaks up into
an exponentially large number of disconnected sectors whose number and
sizes are determined.  The decays of autocorrelation functions in
different sectors follow from an exact mapping to a model of the diffusion
of hard-core random walkers with conserved spins.  These are also verified
numerically. Within each sector the model is reducible to the Heisenberg
model and hence is fully integrable. We discuss additional symmetries of
the equivalent quantum Hamiltonian which relate observables in different
sectors. We also discuss some implications of the existence of an infinity
of conservation laws for a hydrodynamic description. 

\vspace {2cm}

\newpage
\section{Introduction}

The behaviour of an interacting, many-particle system at large length 
scales and times is best described by the evolution equations
for its hydrodynamic modes.  The density fields required
for such a coarse-grained description are determined by the quantities
conserved by the microscopic dynamics.  For example, the
conservation of particle number, momentum and energy in molecular
collisions give rise to the Navier-Stokes equations.
What are the appropriate variables for a hydrodynamic description of a system
with an {\it infinity} of conservation laws ?  This question 
assumes significance in the context of models in which the phase space 
is Many-Sector-Decomposable (MSD), such as the recently investigated
model for the deposition and evaporation of k-mers on
a line \cite{bar1,bar2}.
In this class of models [3-7], the existence of an
infinity of conservation laws implies that the phase space can be 
decomposed into a very large number of disconnected sectors, with the number 
of such sectors typically growing
{\it exponentially} 
with the size of the 
system\cite{vank}. 
As a result, ergodicity is strongly broken, and time-averaged properties
such as correlation functions vary from sector to sector and do not
equal averages over the full phase space.

The best studied model in this class is the trimer deposition-
evaporation (TDE) model\cite{bardha,dhabar}.  The infinity of 
conservation laws in 
this model can equivalently be encoded in
a single conservation law of the so-called ``irreducible string''
\cite{dhabar}.  The long-time decay of the density-density autocorrelation
function was found to differ from sector to sector.
To explain this diversity of dynamical behaviour, it was argued in
\cite{bardha} that the long-time behaviour of autocorrelation functions
in the TDE model is the same as that in a model for diffusing hard-core
random walkers with conserved spin (HCRWCS), the spin sequence of the
random walkers being the same as that of the irreducible string of the TDE
model.  While this conjecture has been very successful in describing
the decay of autocorrelation functions in the TDE model, its precise
justification has not been available so far. 

In this paper, we define a model of diffusing, reconstituting dimers 
on a line which shares many of the features of the TDE
model.  These include the MSD property, conservation laws encoded in
an irreducible string, and different decays of autocorrelation functions
in different sectors, which can be understood using its connection to
the HCRWCS model.  However, unlike the TDE model, the present model is
exactly solvable, and its stochastic matrix can be written as the
Hamiltonian of a quantum spin-$1/2\/$ chain with three-body
interactions. It is closely related to the Bariev model %
\cite{bariev}, another
integrable quantum spin model with three-spin interactions. We show that
our model has an exact equivalence with the
ferromagnetic Heisenberg model.  Interestingly, however, the number of
sites in the original model and the equivalent Heisenberg spin
chain are not equal.  We show that the equivalence to HCRWCS is
{\it exact} for this model, and hence {\it deduce} the
different behaviour of the autocorrelation function in different sectors.
Our model can be viewed as a special case of the
more general symmetric exclusion process with several species of
particles, which has been defined earlier by Boldrighini {\it et  al.}
\cite{boldri}, and has attracted some attention 
recently\cite{derrida,ferrari,schmittmann} .

The quantum spin Hamiltonian corresponding to our model is 
of interest
for two other reasons: firstly, 
in one-dimensional models, integrability is usually established by
constructing a one-parameter family of commuting transfer matrices 
through finding an $R$ matrix which satisfies the Yang-Baxter Equation
\cite{ma90}.  In our model, as also in some other models like the TDE
model, we can construct such a family of nontrivial commuting matrices
{\em without invoking an $R$ matrix}. 
Secondly, the model shows the existence of three different 
sets of conservation laws. 
The number of conservation laws in each set is proportional to the
number of sites, and becomes infinite in the thermodynamic limit. 
The first set consists
of conservation laws which are implied by the conservation of the
irreducible string in our model. 
It is responsible for the decomposition of phase space into an
exponentially large number of sectors.
Further, in each sector, 
corresponding to a
particular irreducible string, the model is equivalent to a Heisenberg
spin chain.  The latter is known to be an integrable model and has an
infinite number of independent constants of motion which commute with
the Hamiltonian.  For each of these observables, there is a
corresponding constant of motion in our dimer diffusion model.
These constants of motion constitute the second infinite set.
The third set of
constants of motion of the model is related to the existence of
a special symmetry in the model, which relates the time evolution in
different sectors.

This paper is organized as follows: In section 2 we define the model.
In section 3, after discussing the nature of the steady states in
the MSD models, we define the irreducible string for our
model and show that it is a constant of motion.  We determine the
number and sizes of sectors into which the phase space breaks up due
to the conservation of the irreducible string. 
In section 
4, we demonstrate the precise equivalence of the model to the 
HCRWCS model and show that it can be viewed as a special case of a generalized 
symmetric exclusion process with several species of particles.  In
Section 
5, we write the transition rate matrix for our model as the Hamiltonian of a 
quantum spin chain, and show that within any specified sector, this 
Hamiltonian is the familiar Heisenberg Hamiltonian, 
known to be completely diagonalizable using the Bethe ansatz.  
In Section 6, we discuss the 
three sets of
conservation laws of the quantum Hamiltonian.
In Section 
7, we use known results 
for the symmetric exclusion process (of one species) on a line to show how
time-dependent correlation functions in any arbitrary sector in our 
model can be determined.  In some selected representative sectors, we have 
explicitly calculated
the behaviour of the time-dependent density-density
autocorrelation function for large times.  These show different
behaviour on different sublattices, and different asymptotic decay laws
($t^{-1/4},~t^{-1/2}$, and sometimes a stretched exponential decay of the
form $\exp [-(t/\tau)^{1/2}]$ for intermediate times).  These predictions 
are verified by Monte Carlo simulations which are discussed in Section %
8. Section 
9 contains a discussion of how the hydrodynamic limit of the
equations of motion for coarse-grained fields is influenced by the 
existence of an infinity of conservation laws.  Finally, Section 
10 contains a summary of our results,
a discussion of their significance, and possible extensions of our analysis
to other models having the MSD property.

\section{Definition of the Model}

The model is defined as follows: We consider a line of $L$ sites.  At
each site $i$ of the chain, we define an integer variable $n_i$ which
takes the value 1 if the site is occupied by a single particle and 0
if it is unoccupied.  A configuration of particles 
is characterized by the set
$\{n_i\}$, which may alternatively be written as an $L$-bit binary
string, e.g.
\be
\ldots 0100010011011110010110 \ldots
\ee
Clearly there are $2^L$ distinct configurations of the system.  The
system undergoes a stochastic evolution under a Markovian dynamics
specified by the following transition rates: In a small time interval
$dt$, any triplet 011 of consecutive sites in a configuration
can change to 110 with probability $\lambda
dt$; it remains unchanged with probability $1-\lambda dt$.  Likewise,
a triplet 110 changes to 011 with
probability $\lambda dt$ and remains unchanged with probability 
$1-\lambda dt$.  These processes may be represented by a
``chemical'' equation
\be
011 \rightleftharpoons 110.
\ee 
By rescaling time, we set $\lambda = 1$.  Thus the model
may be thought of as a lattice gas model, where pairs of particles
(also called dimers) can diffuse, but not
single particles.  However, these pairings are impermanent and the
dimers can `reconstitute'.  For example, in the sequence of transitions
\be
11010 \rightarrow 01110 \rightarrow 01011,
\ee
the middle particle is paired with the particle to the left in the first
transition and with the particle on the right in the second
transition.  Our model is thus a model for diffusing, 
reconstituting dimers (DRD) on a line. 

An alternative description of the DRD dynamics is in terms of assisted
hopping of particles to next-nearest neighbor sites: a particle jumps
two steps to the left or right to an empty site at unit rate, if and 
only if the
intervening site is occupied.  In still another description, it 
may be thought of as hopping of 0's by 2 steps left or right at a 
constant rate, if no other 0's intervene.

It is useful to compare the DRD dynamics with the TDE dynamics.  In the
latter case, the allowed transitions are given by the chemical
equation
\be
111 \rightleftharpoons 000.
\ee %
In this paper unless otherwise stated, we shall assume fixed end
boundary conditions throughout, so that no particles can enter or
leave the system at the ends of the chain.

Higher dimensional generalizations of DRD dynamics are obvious,
but are not very amenable to analytical treatment yet. They will 
be discussed briefly in the last section.

\section{Phase-Space Decomposition and the Irreducible\break String}

In the steady state of a stochastic system with equal transition
probabilities between pairs of configurations,
all configurations accessible to the system occur with equal probability. 
Thus the left eigenvector $(1,1,1,1, \ldots)$,
is also a right eigenvector, and represents a steady state.
Many-sector decomposability
implies the existence of an exponentially large (in the system size)
number of degenerate left (and hence right) eigenvectors and therefore 
an infinitely
large number of steady states in the limit of infinite
system size. In MSD systems, therefore, the right eigenvector
$(1,1,1,1, \ldots)^T$ with all components equal to 1 is best thought 
of as a linear combination of steady states in each sector. 

The dynamics of the DRD model is strongly nonergodic : It is not
possible to reach all the $2^L$ configurations of the $L$-site chain
from any starting configuration using the rules of the dynamics.  For
example, the dynamics clearly conserves the number of particles, and
configurations having different numbers of particles belong to different
sectors.  In fact, it is easy to see that 
in a two-sublattice decomposition of the linear chain, the total numbers
of occupied sites on the odd and even sublattices are separately conserved.  

These, and many more conservation laws, are concisely expressed as the
conservation law of a quantity called the Irreducible String (IS).  The
IS corresponding to a given configuration is defined
as follows: We start with the $L$-bit binary string of 0's and 1's
specifying the configuration.  If there is any pair of adjacent 1's in
the string, both these characters are deleted, reducing the length of
the string by 2.  The procedure is repeated until no further deletions
are possible.  The resulting string is called the IS
corresponding to the configuration.  For example, corresponding to the
configuration $0100101101110011110$, the IS is
$01001001000$.

It is easy to see that we get the same IS for a given
initial string, whatever the choice of
pairs to be deleted and whatever their order of deletion.  
If we get configuration ${\cal C}'$ from
configuration ${\cal C}$ in one elementary step, 
${\cal C}$ and ${\cal C}'$ have the same IS.
Thus the IS is a constant of motion.  

One may note the similarity of the construction of the irreducible
string in our problem with that in the TDE case where
all occurrences of 000 or 111 are recursively deleted.  However, in the
present case, unlike the TDE model, it is sufficient to scan the
configuration from left to right only once to obtain the IS.

We now show that if two configurations ${\cal C}$ and ${\cal C}'$ have the same
IS, one can be changed to the other using the steps of
the DRD dynamics.  Using the dynamical rules, we can push any dimer towards
one end of the string (say the right),
and thus transform any configuration to a {\it standard}
configuration in the sector, in which the configuration is the
IS followed by all the dimers.  As the dynamical steps
are reversible, if ${\cal C}$ and ${\cal C}'$ have the same IS, they can be
transformed into the same standard configuration, and thus into each
other.  This implies that the decomposition of phase space into
disjoint sectors using the conservation law of the IS 
is maximal, and one cannot find additional conservation laws which
will break the sector further into subsectors.  Thus the set of all
possible irreducible strings is in one-to-one correspondence with the
sectors of the phase space and provides a convenient way to label
them uniquely. 

Let $F_n(1)$ and $F_n(0)$ be the number of allowed irreducible strings
of length $n$, whose first character is 1 and 0 respectively.  As
there are no two consecutive 1's in the IS, the
F's satisfy the recursion relations
\begin{eqnarray}
F_n(0) &=& F_{n-1}(0) + F_{n-1} (1), \nonumber \\
F_n(1) &=& F_{n-1} (0).
\end{eqnarray}
With the boundary conditions
\be
F_1(0) = F_1(1) = 1,
\ee
these recursion equations are easily solved. We see that $F_n(1)$
is the $n^{\rm th}$ Fibonacci number.  The total number of distinct
IS's of length $n$ is given by 
\be
N_n = F_n(0) + F_n(1).
\ee
Thus $N_n$ increases as $\mu^n$, for large $n$, where $\mu =
(\sqrt5 + 1)/2$.

For a line of length $L$, the largest possible value of $n$ is $L$.  All
configurations corresponding to $n=L$ cannot evolve to any other
configuration. These are said to be fully jammed and each constitutes
a separate sector having only one configuration.
In a sector with $r$ diffusing dimers, the
length of the IS is $L-2r$.  Summing over $r$, we
get the total number of sectors in our model to be 
\be
N_{\rm total} = \sum^{[L/2]}_{r = 0} N_{L-2r}.
\ee
Thus $N_{\rm total}$ also increases as $\mu^L$, for large $L$.

The definition of the IS is slightly more complicated
for periodic boundary conditions.  It is easily seen that if two
irreducible strings are related to each other by an even number of
cyclic shifts, they have to be considered as equivalent.  Thus for the
configuration $10100111$ on a ring with $L=8$, the irreducible
strings $1000$ and $0010$ have to be considered as equivalent, but
these are different from $0100$ or $0001$.  This 
makes
the counting of sectors for periodic boundary
conditions 
a bit more complicated.

\section{Equivalence to a Model of Hard-Core Random Walkers with
Conserved Spin}

In the algorithm to determine the irreducible string corresponding to
a given configuration, the choice of which pair of adjacent 1's to
delete is immaterial.  One is therefore free to use additional precedence
rules to select which characters in a sequence of consecutive 1's are
deleted.  Let us adopt the convention that we scan the string from left
to right, and delete a pair of 1's whenever they are first encountered.
Using this convention, for any given configuration of the DRD
model, the position of characters which are 
not
deleted is also uniquely
determined.  As the configuration evolves in time, the positions of
these undeleted characters change, but their number and relative
order is conserved.  The elements of the IS may then be viewed as a 
set of interacting hard core particles
(random walkers), which undergo diffusive motion on a line,
but cannot cross each other.  

Conservation of
the IS implies that there is a one-to-one correspondence
between a configuration of the DRD model, and $\{X_i(t)\}$ where $X_i(t)$
denotes the position of the $i^{\rm th}$ unreduced character from the left
(called the $i^{\rm th}$ walker).  The number of walkers equals the
length of the IS in the sector and does not change with
time.  Each walker $i$ carries a `spin' label $S_i$ ($S_i
= 0$ or 1), which is conserved.  More precisely, if in a given sector
the IS is $S_1 S_2 \ldots S_\ell$, and the position of
the $\ell$ walkers are $X_i(t)\/$ with $i = 1~{\rm to}~\ell$, then
correspondingly, in the DRD model, the configuration $\{n_i\}$ is given
by
\be
\begin{array}{l}
n_j = S_i,~ {\rm~if~the~} i^{th} {\rm ~walker~is~at~site~} j \nonumber\\
~~~~ =1,~ {\rm~~if~no~walker~is~at~site~} j.
\end{array}
\ee
This establishes the equivalence of the DRD model with the model of
hard-core random walkers with conserved spin (HCRWCS).

The evolution rules for this equivalent HCRWCS model are easily
written down, and are seen to be a special case of the general
$k$-type exclusion process introduced first by Boldrighini {\it et al.} 
\cite{boldri}.
This exclusion process (XP) is defined in the following way. We consider a
lattice in which each site is occupied by one of $k$ different
types of particles.  Particles can interchange positions with other
particles at neighboring sites at a rate which depends on the types of
particles interchanged.  In general, the rate at which particles of
type $m$ and type $n$ change position is $\gamma_{mn} = \gamma_{nm}$.
In this model it is assumed that $\gamma_{mn}$ is only a function of
$min(m,n)$.  The case with only 2 types of particles is the ordinary
symmetric exclusion process, where the two types are usually called
particles and vacancies.  Asymmetric versions of this model,
where $\gamma_{mn} \not= \gamma_{nm}$,  have also been studied.

The DRD model corresponds to a special 3-type exclusion process. This
is seen as follows:  In the HCRWCS model, if there is a walker with
conserved spin label 1 (say at site $X_i$) then there must also be a random
walker with spin label 0 at site $X_i+1$. Therefore
\be
X_{i+1} = X_i + 1.
\ee
Thus the $i^{\rm th}$ and $(i+1)^{\rm th}$ walker always move
together if the $i^{\rm th}$ walker 
has the spin label
1.  It is thus advantageous to think 
of these two walkers as a single
walker.  But then there are two kinds of walkers: `double' walkers of the
10 type which occupy two adjacent sites (we shall call these particles of
type B); and single site occupying
walkers (isolated vacancies, to be called type C particles).  All pairs of
sites not occupied 
by particles in the HCRWCS picture may be said to be occupied by particles 
of type A, which also occupy two adjacent sites %
each. 
These are the dimers which 
are deleted in the construction of the IS.

A configuration of this 3-type exclusion process is then given by a
string of characters of the kind $\ldots ABCBAABCBB \ldots$.  The
corresponding unique DRD configuration is determined by the direct
substitution $A \rightarrow 11, B \rightarrow 10$ and $C \rightarrow
0$.  It is easy to see that DRD dynamics corresponds to the rule that
type A particles can exchange positions with type B or type C
neighbors with equal rate. Type B and C particles cannot exchange
positions, and thus their relative order is preserved by the
dynamics.

With periodic boundary conditions, all DRD configurations 
can be decomposed uniquely into A, B and C type particle configurations.  
With free boundary conditions it is possible to get a single unpaired 
1 at the right end, which cannot be combined with a 1 or 0 to give 
rise to an A or B type particle.  This can be taken care of by choosing a
boundary condition in which the last site on the chain is always 
0, which does not evolve.

The conservation law of the IS in this language is the
simple statement that in a given configuration specified by a string
composed of characters A, B or C, deleting all occurrences of A's
leaves us with an invariant string which specifies the relative order
of B and C type particles in the initial state. The MSD property of the
DRD model follows simply from this property.

Using this picture of a 3-type exclusion process, it is straightforward
to determine the number of configurations which constitute a given
particular sector.  Consider free boundary conditions for convenience.
In a sector in which there are $N_A$ particles of type A, $N_B$
particles of type B, and $N_C$ particles of type C, the total number
of distinct configurations with relative orders of B and C specified is
the same as the number of configurations of $N_A$ type A particles 
and $(N_B + N_C)$
non-A particles in a 2-type exclusion process.  Denoting this number by
$\Omega^{\rm free}$ $(N_A, N_B, N_C)$ we have 
\be
\Omega^{\rm free}(N_A, N_B, N_C) = (N_A + N_B + N_C)!/N_A!(N_B + N_C)!
\ee
The total size of the lattice in the DRD model is $L=2 N_A + 2N_B + N_C$
and the number of zeros is $N_B + N_C$.

\section{Quantum Spin Hamiltonian Corresponding to the Rate Matrix}

It is straightforward to write down the relaxation matrix as the
Hamiltonian of a quantum spin chain \cite{alc}.
Let $P({\cal C},t)$ be
the probability that a classical system undergoing Markovian evolution
is in the configuration ${\cal C}$ at time $t$.  These probabilities 
evolve in time
according to the master equation
\be
{\partial P ({\cal C},t) \over \partial t} = \sum_{{\cal C}'} 
\left[-W ({\cal C}
\rightarrow {\cal C}')\, P({\cal C},t) + W({\cal C}' 
\rightarrow {\cal C})\, P({\cal C}',t)\right]
\ee
where the summation over ${\cal C}'$ is over all possible 
configurations of the system
and $W({\cal C} \rightarrow {\cal C}')$ is the transition rate 
from configuration ${\cal C}$ to
configuration ${\cal C}'$.

We 
construct 
a Hilbert space, spanned by basis vectors $|{\cal C}\rangle$,
which are in one to one correspondence with the configurations
${\cal C}$ of the system.  A state with probability weight 
$P({\cal C},t)$ %
for the configuration $C$
is represented in this space by a vector
\be
|P(t)\rangle = \sum_{{\cal C}} P ({\cal C},t) | {\cal C}\rangle
\ee
The master equation can then be written as
\be
{\partial |P(t)\rangle \over \partial t} = - {\hat H} |P(t)\rangle
\ee
This equation can be viewed as an imaginary-time Schodinger equation
for the evolution of the state vector $|P(t)\rangle$ under the action
of the quantum Hamiltonian ${\hat H}$.

The $2^L$ configurations of the DRD model on a line are in one-to-one
correspondence with the configurations of a spin-$1/2$ chain of $L$
sites.  At the site $i$, the spin variable is $\sigma_i^z = 2n_i - 1$
taking values $+1$ and $-1$.  It is
then straightforward to write ${\hat H}$ as the Hamiltonian of the spin
chain in terms of the Pauli spin matrices $\sigma_i$.  We find that
${\hat H}$ has local 2-spin and 3-spin couplings, and is given by
\be
{\hat H} = -\frac{1}{4}\sum_i \left[\vec\sigma_{i-1} 
\cdot \vec\sigma_{i+1} - 1\right] \, (1+\sigma^z_i)
\ee
The relationship of this Hamiltonian to the Hamiltonian of an integrable
spin model with three-spin couplings proposed and solved by Bariev\cite{bariev},
will be discussed in the final section.

It is useful to write this Hamiltonian in terms of fermion operators
$c^+_i$ and $c_i$ defined by the %
standard Jordan-Wigner transformation.
In terms of these fermion operators, ${\hat H}$ can be written as
\be
{\hat H} = \sum_i \left[c^+_{i-1} n_i c_{i+1} + c^+_{i+1} n_i c_{i-1} + 
n_i (n_{i-1} - n_{i+1})^2\right]
\ee
where $n_i = c^+_ic_i$ is the number operator at site $i$.  The first
two terms of this Hamiltonian represent assisted hopping over an
occupied site, and the last term describes two- and three- body
potential interaction between nearby sites.  Similar hopping terms are
encountered in a model studied by Hirsch in the context of hole
superconductivity\cite{hirsch}.

\section{Conservation Laws for the DRD Model}

In the previous section, we have seen that the evolution of
probabilities in a classical Markov process can be cast as the
evolution of the wavefunctions of a suitably defined quantum mechanical
problem.  However, the concept of constants of motion has somewhat
different meanings in these two cases.  This is explained below.

An observable ${\cal O}$ of the classical stochastic process is a function
which assigns a real number ${\cal O}(C)$ to each possible 
configuration $C$ of
the system.  We say that ${\cal O}$ is a classical constant of motion if its
value ${\cal O}(C)$ does not change, even as the configuration changes
with time.  In a quantum mechanical formulation, observables are
represented by matrices $\hat{\cal O}$, and the quantities of interest are
the expectation values at time $t$ obtained through the formula
\be
\langle \hat {\cal O} \rangle_t = \langle 1 | \hat {\cal O} | P(t) \rangle
\ee
where $\langle 1|$ is the row vector with all entries 1.  We say that
${\cal O}$ is a quantum mechanical constant of motion if $\hat {\cal
O}$ commutes
with the Hamiltonian $\hat H$.  In this case, it is easy to see that
the expectation value in Eq. (6.1) does not change in time. [Note that
we are using a different prescription for evaluating
expectation values than in standard quantum mechanics.]

Clearly the class of all possible quantum mechanical constants of
motion is much larger than that of classical constants of motion.  The
latter corresponds to diagonal matrices in the natural basis of the
system, i.e. that given by basis vectors $|C\rangle$.  The quantum
mechanical operators are in general not diagonal in this basis.

In the DRD model, both kinds of constants of motion are
found.  To be specific, we find three different classes of constants
of motion.  We discuss them below. 

\medskip

\noindent \underbar{A. The classical constants of motion}

\smallskip

Let us consider two matrices $A(0)$ and $A(1)$, such that 
\be
A(1) A(1) A(0) = A(0) A(1) A(1)
\ee
For any configuration $C$ of the DRD model, specified by occupation
number $\{n_i\}$, we associate a matrix $I_L$ given by
\be
I_L = \prod^L_{j=1} A(n_j),
\ee
where the matrix product is ordered from left to right in order
of increasing $j$.

As $A(1)^2$ commutes with $A(0)$, it follows that the matrix elements of
$I_L$ are classical constants of motion of the DRD model.  We may
choose $A(0)$ and $A(1)$ to be $2 \times 2$ matrices.  Then Eq. (6.2)
does not determine the matrices completely as these constitute only 4
equations for 8 variables.  Thus, we can choose $A(0)$ and $A(1)$ to
depend on a real parameter $\lambda$, satisfying Eq. (6.2) for all
values of $\lambda$.  A simple choice satisfying Eq. (6.2) is
\be
A(0) = \left[\matrix{1-\lambda & \lambda \cr 1 & -1}\right];
\,\,\,\,\, A(1) = \left[\matrix{1 & 0 \cr 0 & -1}\right]
\ee
It is easy to see that specification of $I_L(\lambda)$ specifies the
irreducible string and that the correspondence between these is one to one.
We thus introduce the $(2 \times 2)$ matrix-valued classical operator
(it is diagonal in the configuration basis)
\be
\hat I_L(\lambda) = \prod^L_{j = 1} A(\hat n_j, \lambda)
\ee
where $\hat n_j = \sigma^+_j \sigma^-_j$ is the number operator at
site $j$.  Then $\hat I_L$ commutes with $\hat H$, and is a classical
constant of motion.  Since $\hat I_L$ is a polynomial in $\lambda$, we
can write it as
\be
\hat I_L (\lambda) = \sum^L_{r=0} \hat Q_r \lambda^r
\ee
where $\hat Q_r$ are $2 \times 2$ matrices, each element of which is a
classical constant of motion of the DRD model.  In the limit of $L
\rightarrow \infty$, we get an infinite number of classical commuting
operators $\{Q_r\}$ which also commute with the Hamiltonian $\hat H$. 

In fact, if two configurations have the same values of $\hat Q$ for all
$Q$, then they have the same irreducible string, and must belong to
the same sector.  As the IS provides the maximal decomposition of phase
space, there can be no other independent classical constants of motion
in our model. 

\medskip

\noindent\underbar{B. The quantum-mechanical constants of motion in a
fixed sector}

\smallskip

Since ${\hat H}$ commutes with ${\hat I}_L(\lambda)$, it has a
block-diagonal structure, with each block corresponding to a sector of
the phase space, and to a particular IS.  The
task of diagonalizing ${\hat H}$ then reduces to that of diagonalizing it in
each of the sectors separately.
Consider any one of these sectors.  Let the IS ${\cal I}$ for
this sector, in the 3-species exclusion process notation, be a string of
$q \, B$'s and $r\, C$'s in some order, say ${\cal I} = BCBBCBCC \ldots$
The number of diffusing dimers in this sector then is $p= (L-2q-r)/2$.  A
typical configuration in this sector is specified by a string of
length $(p+q+r)$ with $p\, A$'s interspersed between the characters of
the IS, e.g. BCABAABCBCAC $\ldots$.

We define a chain of spin-1/2 quantum spins ${\tau_i}$, with $i$ ranging
from $1\/$ to
$L_\tau = (p+q+r)$.  For each configuration in the sector ${\cal I}$, we
define a corresponding configuration of the $\tau$-spin chain by the
rule that $\tau_j^z = +1$ if the $j^{\rm th}$ character in the string
specifying the configuration is $A$.  For $\tau^z_j = -1$, 
the corresponding
character can be either $B$ or $C$.  But this degeneracy is completely
removed by using the known order of these elements in the irreducible
string ${\cal I}$.  Thus there is a one-to-one correspondence between the
configurations of the DRD model in the sector ${\cal I}$, and the
configurations of the $\tau$-chain with exactly $p$ spins up.

The action of the Hamiltonian $\hat H$ on the subspace of configurations
in the sector ${\cal I}$ looks much simpler in terms of the $\tau$-spin
variables.  We have already noted, in Section IV, that the evolution 
in terms of the $\tau$-variables is that of a simple exclusion process. 
The quantum Hamiltonian for this process is the well known Heisenberg
Hamiltonian
\be
{\hat H}_{Heis} = - \sum^{L_\tau -1 }_{i=1} \left(\vec \tau_i \cdot \vec
\tau_{i+1} - 1 \right).
\ee
Note that ${\hat H}_{Heis}\/$ looks different for sectors
with differing $\L_\tau$'s.

We would now like to construct a one-parameter family of operators
that commute with ${\hat H}_{Heis}$.  This is a well-known construction for the
Heisenberg model \cite{panchgani}.  We use periodic boundary conditions for
convenience.  Define $2 \times 2$ matrices $L_j(\mu)$ whose elements
are operators acting on the spin $\tau_j$, and $\mu$ is a parameter
\be
L_j(\mu) = \Biggl( \matrix{1 + i\tau_j^z \mu & i\mu \tau^-_j \cr i\mu
\tau_j^+ & 1 - i\tau^z_j\mu} \Biggr),
\ee
and define
\be
\hat \T(\mu) = Tr \prod^{L_\tau}_{j=1} L_j(\mu).
\ee
Then it can be shown that \cite{panchgani} for all $\mu,\mu'$
\be
\left[\hat \T (\mu), \hat \T(\mu')\right] = 0.
\ee
Writing $\ln \hat\T(\mu) = \sum^\infty_{r=1} \mu^r \hat J_r$
we get $\left[\hat J_r, \hat J_s\right] = 0$, for all $r,s$ and $\hat
J_2 = {\hat H_{Heis}}$.  Thus the set of operators $\{\hat J\}$ constitute a
set of quantum mechanical constants of motion for the Hamiltonian
${\hat H}_{Heis}\/$ in each sector separately, and hence for ${\hat H}\/$.

Note that for all these operators $\{\hat J\}$, the corresponding
matrices in the configuration basis have a block diagonal structure,
where the blocks correspond to different sectors, but are off-diagonal
within a block.  

\medskip

\noindent\underbar{C. The Inter-sector Quantum Mechanical Constants of
Motion}  

\smallskip

The Hamiltonian ${\hat H}$ has still another additional infinite set of
constants of motion.  These are related to the existence of an
additional symmetry in the model.  Clearly, replacing a $B$-type
particle by a $C$-type particle does not affect the dynamics of the
exclusion process.  Thus, the full spectrum of eigenvalues of ${\hat H}$
in any two sectors of the DRD model with different irreducible
strings, but having the same values of $p$ and $(q+r)$, is exactly the
same. Such a symmetry may be viewed as a local gauge symmetry between
the `$B$' and `$C$' ``colors''.  Changing the character in the
IS from $B$ to $C$ (or vice versa) changes the length
of $L$ by 1.  This symmetry therefore relates two sectors of the DRD 
model with different sizes of the system. 

The simplest inter-sector operators which preserve the
total length of the chain $L$ are operators which interchange two
characters of the IS.
Let ${\hat K}_j$ be the operator which interchanges the $j^{\rm th}$
and $(j+1)^{\rm th}$ characters of the IS.  Then
clearly, we have
\be
\left[{\hat H},{\hat K}_j\right] = 0 \,\, {\rm for}\,\, j = 1 \,\, {\rm
to} \,\, L_\tau - 1.
\ee

For a fixed value of $q$ and $r$, there are $(q + r)!/(q!r!)$
different irreducible strings, and as many different sectors.  The
`color' symmetry implies that the spectrum of ${\hat H}$ is the same in
each of these sectors.

We have not attempted to write down explicit expressions for the
operators ${\hat J}_j$ and ${\hat K}_j$ in terms of the original variables
$\{\sigma_k\}$.  This seems quite difficult, as the transformation
between these variables is highly nonlinear 
(though easy to implement as an algorithm), and it does not
seem to be particularly instructive at this stage.

\section{Time-dependent Correlation Functions}

The autocorrelation function of the DRD model shows interesting 
variations from one sector to another. Such variations occur despite the
fact that, in each sector, there is a mapping between the DRD model and
the simple exclusion process whose dynamics is known to be governed
by diffusion. As we will see below, this mapping leads to a 
correspondence between tagged hole correlation functions %
(defined below) 
in the two  problems, but the form of autocorrelation function decays can be
quite different.

Consider a particular sector with IS ${\cal I}$. Let the number
of diffusing A, B and C particles in the equivalent 3-species exclusion
process be $p,q,$ and $r\/$ respectively. %
A hole (vacant site) is associated with a $B$ or $C$ particle.
In the IS, let $b(k)$ be the number
of B's to the left of the %
$k$-th hole. Evidently, the function $b(k)$
specifies the IS completely. 

Different configurations in the sector are obtained from different 
distributions of A's in the background of B's and C's. If there are
$a_k$ A's to the left of the k'th hole, the location of this
hole in the DRD and XP problems is
\begin{eqnarray}
y_{XP}(k,t) &=& k + a_k(t), \\   
y_{DRD}(k,t) &=& k + b(k) + 2 a_k(t).
\end{eqnarray}
Notice that $b(k)$ (unlike $a_k$) is time-independent. Defining a
tagged-hole correlation function for the XP and DRD problems in 
analogy with the conventional tagged-particle correlation 
function\cite{majbar} through
\begin{equation}
\sigma^2(t) = <[y(k,t) - y(k,0)]^2>,
\end{equation}
we see that the simple relationship
\begin{equation}
\sigma_{DRD}(t) =  2 \sigma_{XP}(t)
\end{equation}
holds.  However, no simple, exact equivalence between the two problems 
can be established for tagged-{\it particle} correlations or single-site
autocorrelation functions. This is because the
transformation between spatial coordinates in the DRD and related 
XP problems is nonlinear; a fixed site in the former problem corresponds to
a site whose position changes with time in the latter.

Let $x\/$ and $\xi\/$ denote spatial locations in the DRD and XP
problems respectively and let $a_\xi$ be the number of A's to the 
left of $\xi$.  Evidently, the number of holes (B's and C's) to
the left of this site is $(\xi - a_\xi)\/$. In the equivalent
configuration in the DRD problem, each  A corresponds to two
particles, while the $(\xi -  a_\xi)\/$ B's and C's occupy a length
$b(\xi - a_\xi) + \xi - a_\xi\/$ which depends on the IS in question. 
Thus
\begin{equation}
x = a_\xi + b(\xi - a_\xi) + \xi,
\end{equation}
while the number of particles $a_x$ to the left of $x$ is given
by 
\begin{equation}
a_x =  2 a_\xi + b(\xi - a_\xi).
\end{equation}
The transformation between the integrated particle densities 
$a_x$ and $a_\xi$ is therefore quite complicated. It depends on the IS
through the function $b\/$ and is highly non-linear.

The correlation functions involving $a_\xi$ are quite simple. The
density-density correlation function for the XP in steady state is defined as
\begin{equation}
C_{XP}(\xi,t) = <n^\prime(\xi_0+\xi,t_0+t)n^\prime(\xi_o,t_0)>.
\end{equation}
where $n^\prime$ is a particle occupation number and an average over
$\xi_0$ and $t_0$ is implicit. $C_{XP}$ satisfies the simple
diffusion equation.
\begin{equation}
\frac{\partial C_{XP}(\xi,t)}{\partial t} = \nabla^2_\xi C_{XP}(\xi,t),
\end{equation}
where $\nabla^2_\xi$ is the discrete second-difference operator. For large $t$,
therefore, $C_{XP}$ decays as $t^{-1/2}$. Since $a_\xi$ is a space integral
over $n^\prime(\xi)$, correlation functions involving $a_\xi$'s can
be obtained as well. However, the change of variables from $a_\xi$ to
$a_x$ (Eqs. 7.5 and 7.6) is difficult to perform explicitly.
Nevertheless, we can determine the asymptotic behaviour of correlation
functions $C_{DRD}(x,t)$, defined analogously to Eq. 7.7.
This is illustrated below for various sectors.

The simplest sector is the one characterized
by the IS $101010 \ldots$ of length $\ell$
which is a finite fraction of the total length $L$.  In this sector, 
all the odd sites
are always occupied and the dynamics on the even sites is that of the
simple exclusion process.  Thus $C_{DRD}(0,t)$ 
is zero for $x_0$ on the odd sublattice, while it decays as $t^{-1/2}$ on
the even sublattice.

Next consider the case where the IS consists of
zeros $0000\ldots$, and is of a length $\ell$ which is a nonzero fraction of
$L$. The $t^{-1/2}$ decay of $C_{XP}$  then implies a similar behaviour 
for $C_{DRD}$.

Now consider the general case of an IS whose
$j^{\rm th}$ character is $\alpha_j = 0,1$.  Let us assume that at time
$t_0$, the site $x_0$ is occupied by a particular character of the IS. 
Between the time $t_0$ and $t_0+t$,  let the net number of dimers which cross
the point $x_0$ towards the right be $m$ (a leftward crossing being counted 
as a contribution $-1$ to $m$).  For large times $t$, it
is known that the distribution of $m$ is approximated by a Gaussian
whose width increases as $t^{1/4}$, i.e.
\be 
{\rm Prob} (m|t) \approx {1 \over \sqrt{2\pi}\Delta_t} \exp
\left[{-m^2 \over 2\Delta^2_t}\right].
\ee
where $\Delta_t$ increases as $t^{1/4}$ for large $t$\cite{majbar}.  
If $m$ dimers
move to the right, a site occupied by $\alpha_j$ at time $t_0$ will now be
occupied by $\alpha_{j+2m}$ at time $t_0+t$.  Hence the autocorrelation function
$C(t)$ is approximated by
\be
C(t) = \sum^\infty_{m=-\infty} \overline{\alpha_j\alpha_{j+2m}}
\cdot {\rm Prob} (m|t),
\ee
where $\overline{\alpha_j\alpha_{j+2m}}$ is the average value of the
correlations of characters in the IS averaged over
$j$.  This can have different values for even and odd $j$'s as
a particular element of the IS always stays on one
sublattice.  Define
$\gamma_{\buildrel {\rm odd} \over {\rm even}}  (m) = 
\overline{\alpha_j\alpha_{j+2m}}$, averaged over odd/even sites.  
We have
\be
C_{{\buildrel {\rm odd} \over {\rm even}}} (\tau) = \sum_m
\gamma_{{\buildrel {\rm odd} \over {\rm even}}} (m) \, {\rm Prob}
(m|t).
\ee
If $\gamma (m)$ is a rapidly decreasing function of $m$, then only small
values of $m$ contribute to $C(t)$.  
Thus $C_{DRD}$ varies as $t^{-1/4}$ whenever correlations in the
IS are short ranged. 
In the general case, we can have $C_{DRD}(t)\/$ decaying as 
$t^{-\alpha}\/$, with $0.25 \leq \alpha\/$ by generating 
irreducible strings with $\gamma (m)\/$ such that its Fourier transform
${\tilde \gamma}(k)\/$ varies as $k^{4\alpha-1}\/$ as 
$|k| \rightarrow 0\/$, in such a way that 
Eq. 7.11 yields the desired power of $C_{DRD}\/$. 

Finally consider the sector with the periodic IS $100010001000\ldots$. 
In this case only the even sublattice has density fluctuations, and so
on this sublattice $C_{DRD}$ decays as $t^{-1/2}$.  On the
odd sublattice, $\gamma_m$ has a term proportional to $(-1)^m$ 
which leads to a contribution $\exp (-\sqrt{t/\tau})$  in $C_{DRD}$.
However,  the $t^{-1/2}$ diffusive tail arising from the density
correlation fluctuations dominates at very large times.

\section{Numerical Results}

We have tested the predictions of the previous section by extensive
numerical simulations.  Our simulations were performed on lattices of
size $L=99996$, using periodic boundary conditions.  The initial
configuration with a given IS was generated by
different methods.  The basic update step is as follows: We choose a
site at random out of the $L$ sites. If this site is empty, another is
chosen.  If it is occupied, its two neighbors are interchanged.  If
the neighbors are both empty or both occupied, this does not change
the configuration.  if they are different, the effect is to move a
dimer one unit to the left or right.  A sequence of $L$ such updates
defines a single Monte Carlo step (MCS).

In our simulations, we typically allowed the initial configuration to
evolve over $2 \times 10^4$ MCS before collecting data, which was done
at interval of 10 MCS.  Autocorrelations were separately computed for
even and odd sublattices, averaged over 100 - 500 different
histories, and over all sites of a sublattice. The calculated 
autocorrelation functions were further binned to reduce scatter in 
the data.

\begin{figure}[htb]
\centerline{
        \epsfxsize=12.0cm
        \epsfysize=10.0cm
        \epsfbox{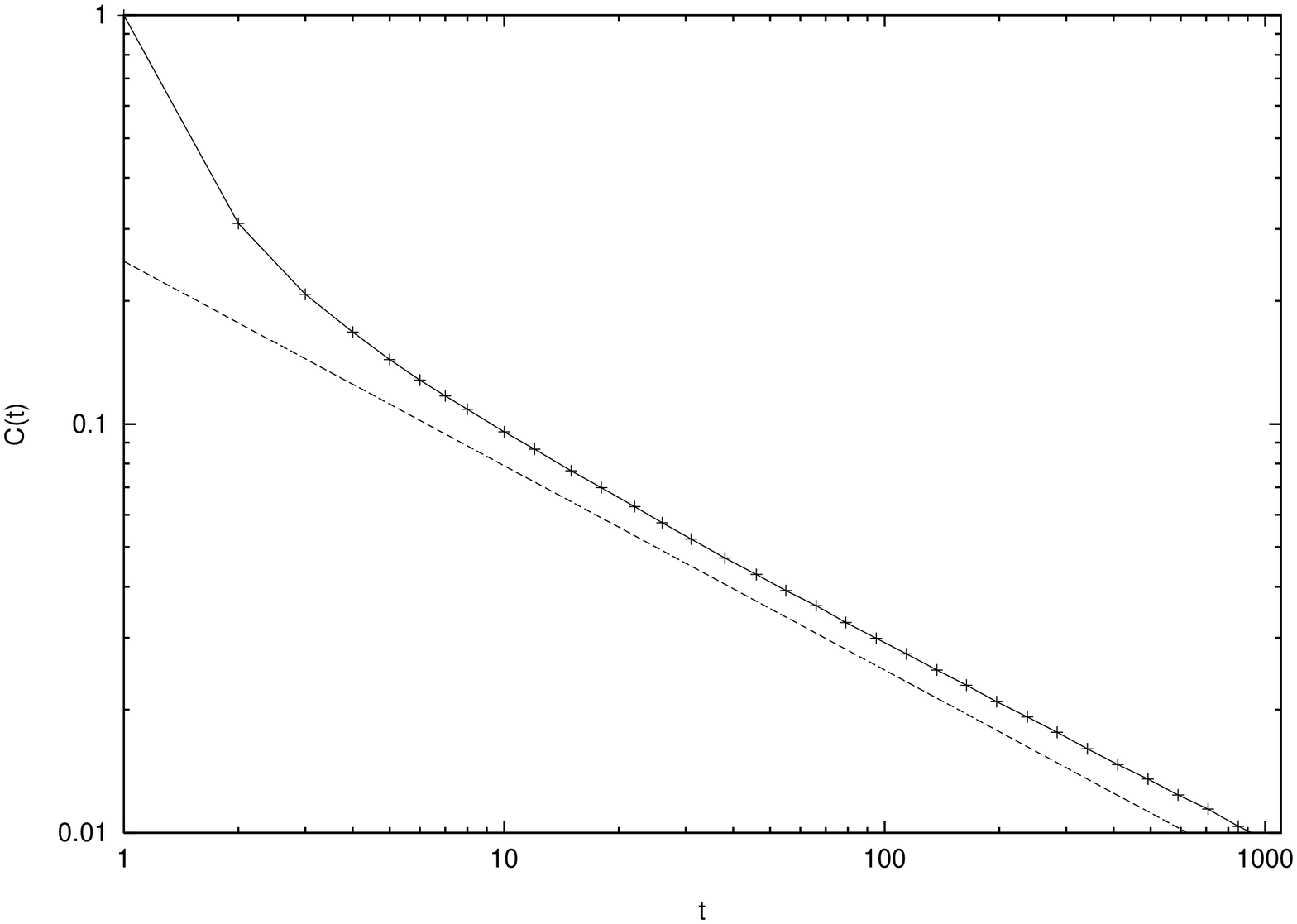}
}
\caption{Autocorrelation function $C(t)\/$ vs. $t\/$ for the sector 
with irreducible string $\ldots 10101010 \ldots \/$. 
The straight line of slope $-1/2$ is a guide to the eye.}
\label{fig:1}
\end{figure}

\begin{figure}[htb]
\centerline{
        \epsfxsize=12.0cm
        \epsfysize=10.0cm
        \epsfbox{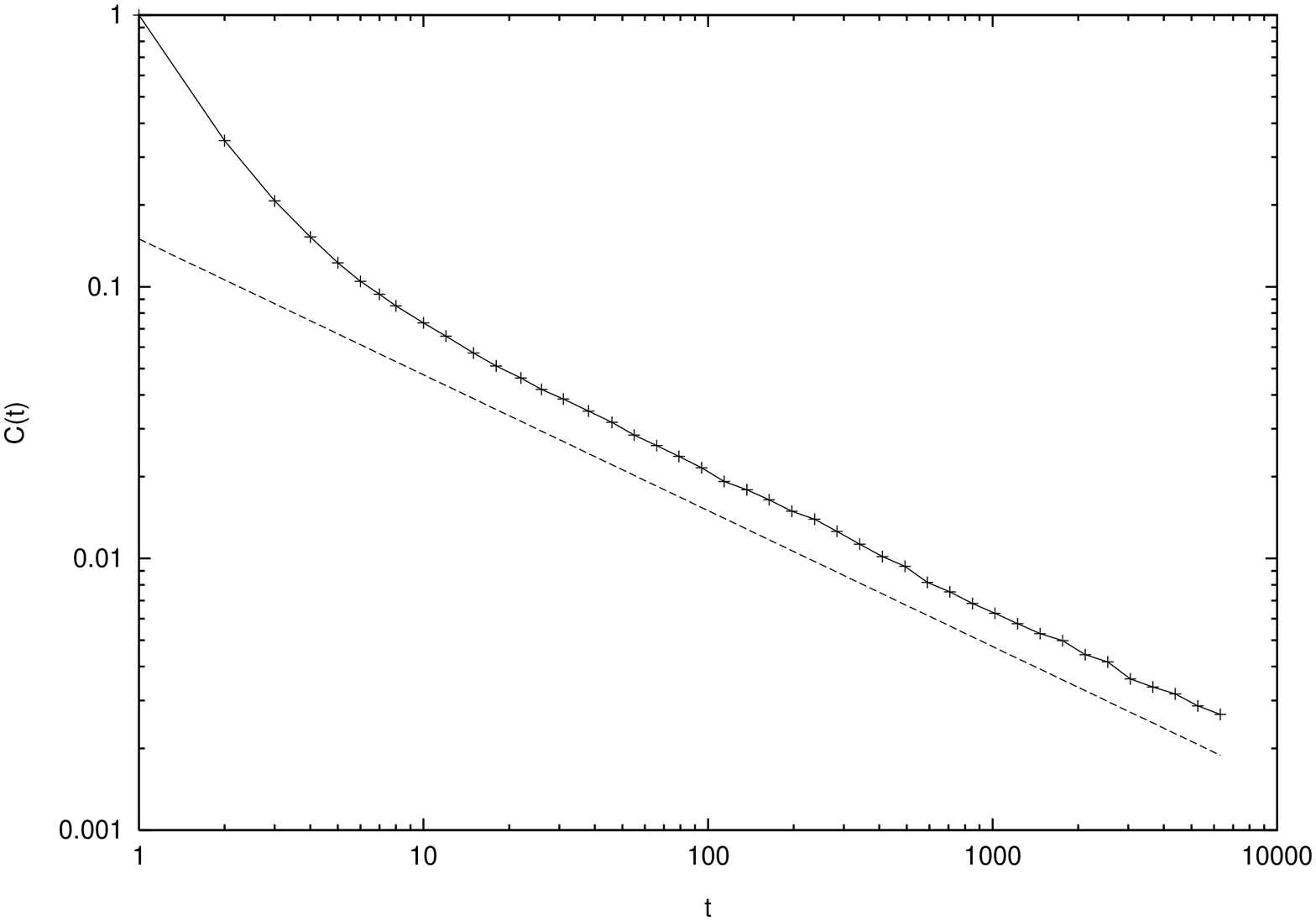}
}
\caption{Autocorrelation function $C(t)\/$ vs. $t\/$ for the sector 
with irreducible string $\ldots 0000 \ldots \/$.
The straight line of slope $-1/2$ is a guide to the eye.}
\label{fig:2}
\end{figure}
\begin{figure}[htb]
\centerline{
        \epsfxsize=12.0cm
        \epsfysize=10.0cm
        \epsfbox{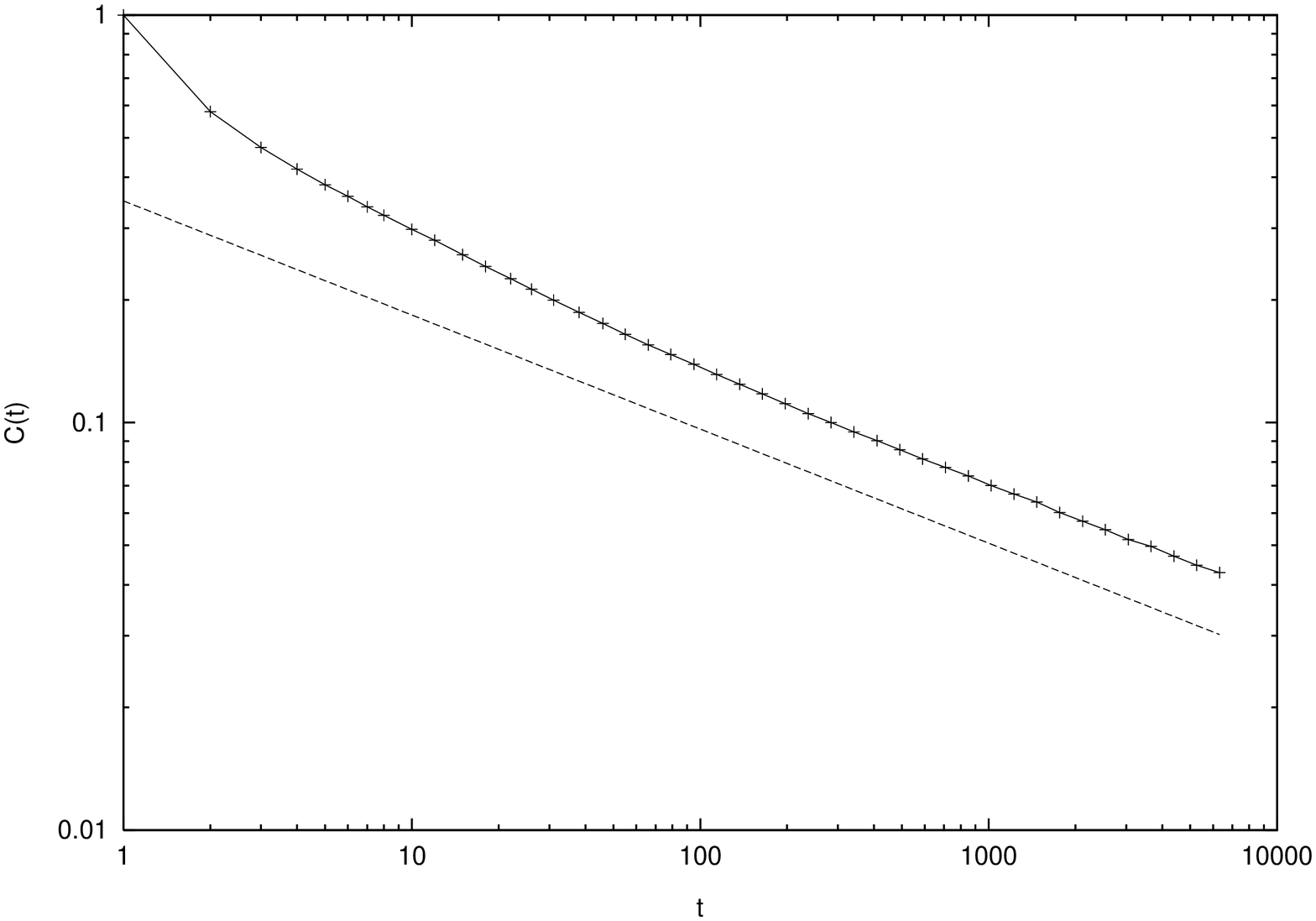}
}
\caption{Autocorrelation function $C(t)\/$ vs. $t\/$ for the random 
sector. The straight line is a guide to the eye and has slope $-0.28\/$.}
\label{fig:3}
\end{figure}
\begin{figure}[htb]
\centerline{
        \epsfxsize=12.0cm
        \epsfysize=10.0cm
        \epsfbox{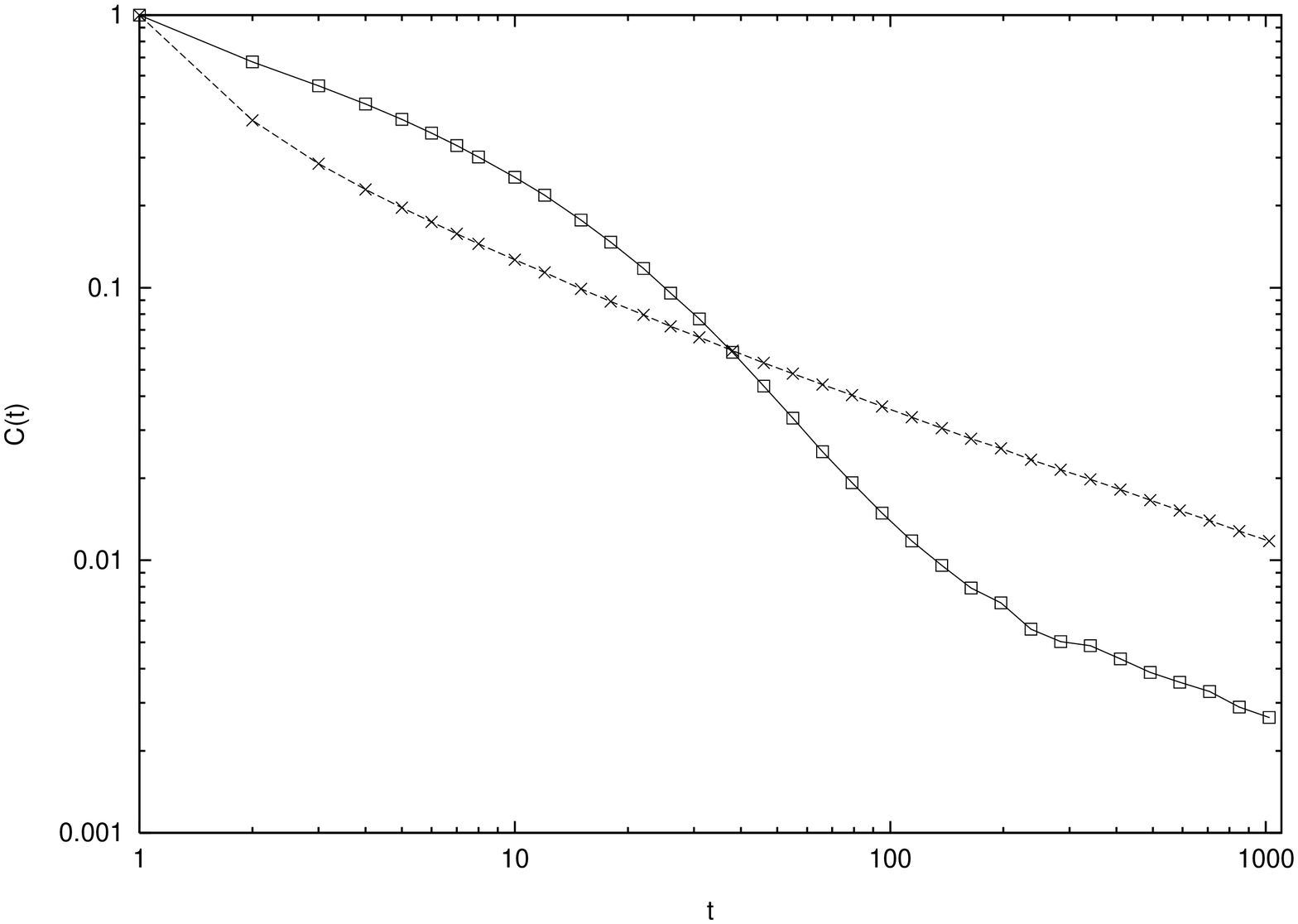}
}
\caption{Autocorrelation function $C(t)\/$ vs. $t\/$ for the sector 
with irreducible string $\ldots 100010000 \ldots \/$. The two curves
are for the different sublattices A and B (see text).}
\label{fig:4}
\end{figure}

In Fig. 1, we show the normalized autocorrelation function $C(t)$ in
the sector where the IS is $101010\ldots$ of length
0.4L.  In this sector, the odd sites are always occupied, and thus the
autocorrelation function on odd sites is trivial.  Our data for $C_{\rm
even} (t)$ shows a clear $t^{-1/2}$ decay for times $t \gsim 20 MCS$.
This is in perfect agreement with the expected diffusive behaviour in
this sector. 

In Fig. 2, we show the results for the sector where the IS
is $0000 \ldots$  To generate the initial configuration in
this case, we used the fact that in the equivalent exclusion process
involving only dimers and vacancies (A and C particles), the steady
state satisfies product measure.  For determining the
autocorrelation function, rather than taking time averages over the
evolution of a single initial configuration, we found it more
efficient to average over an ensemble of different initial conditions
generated with the steady state distribution.  We see a diffusive
behaviour $C(t) \sim t^{-1/2}$ for both the even and odd sublattices,
in agreement with the results of section 8. 

In Fig. 3, we display the results for a randomly generated initial
configuration with equal numbers of 0's and 1's.  The initial
configurations for this sector were generated by adding $L/2$ 1's to
an initially empty lattice to ensure exact 1/2 filling for all
configurations.  Numerical fits to the data show that $C_\alpha (t)
\sim t^{-.28}$ for large $t$, through the convergence to this value is
slow.  This is consistent with the theoretically expected $t^{-1/4}$
behaviour. 

In Fig. 4, we show the results for the IS 
$100010001000\ldots$  The data are consistent with the predictions
stretched exponential relaxation of the form $\exp \left(-
\sqrt{t/\tau}\right)$ for intermediate timescales ($10 < t < 1000$) 
on the odd sublattice, and diffusive $t^{-1/2}$
relaxation on both sublattices at large times. 

\section{Hydrodynamical Description of Many Sector Decomposable 
Systems}

The present study allows us to answer, at least partially, the
question posed in the introduction concerning the hydrodynamical
description of a system with an infinity of conservation laws. 

In the DRD model, such a hydrodynamical description would be in terms
of coarse-grained density fields $\rho_1(x,t)$ and $\rho_2(x,t)$, the
local densities of zeros on the odd and even sublattices
respectively, where $x$ is a continuous variable $(0 \leq x \leq L)$.
Equivalently, we may use the integrated density fields $m_1(x,t)$ and
$m_2(x,t)$, where
\be
m_1(x,t) = \int^x_0 dx' \rho_1(x',t)
\ee
and a similar equation for $m_2(x,t)$.

To write down the evolution equations, we first have to specify the IS
corresponding to the sector.  The hydrodynamical part of the information
contained in the IS may be specified by two functions $\ell_1(m)$ and
$\ell_2(m)$.  These gives the positions of the $m^{\rm th}$ odd- and
even-zero on the IS counting from left [we call a zero an even or
odd zero depending on  whether it is at an even or odd position on 
the IS].  As $\ell$ and $m$ are both much larger than 1, we think 
of $\ell_1(m)$ and $\ell_2(m)$ as real-valued monotonic increasing 
functions of a real argument.  From the definition, it is clear that
\be
\ell_1(0) = \ell_2 (0) = 0
\ee
and
\be
d\ell_1/dm \geq 2, \,\, d\ell_2/dm \geq 2.
\ee
The fact that elements of the IS cannot cross each other immediately
implies that the fields $m_1$ and $m_2$ satisfy the constraint
equation
\be
\ell_1(m_1(x,t)) = \ell_2 (m_2(x,t)), \,\, {\rm for~all} \, x,t.
\ee
The coordinate $\xi$ of the $XP$ is given in terms of $x$ by the
equation
\be
\xi = \left[x - \ell_1 (m_1(x,t))\right]/2 + m_1(x,t) + m_2(x,t)
\ee
For a fixed $t$, on increasing $x$ by a small amount $\Delta x$, the
corresponding increases in $\xi, m_1$ and $m_2$ are given by
\be
\Delta\xi = {1 \over 2} \, \Delta x + \Delta m_2 + \Delta m_1 \left[1
-{1 \over 2}\ell'_1 (m_1)\right]
\ee
where the prime denotes differentiation.
The density of zeros in the exclusion process $\rho_{_{XP}}(\xi,t)$ is
given by $(\Delta m_1 + \Delta m_2)/\Delta \xi$.  From Eq. (9.6), we
get
\be
\rho_{_{XP}}(\xi,t) = (\rho_1 + \rho_2) / \left[{1 \over 2} + \rho_1 +
\rho_2 - {\rho_1 \over 2}\, \ell'_1 (m_1)\right]
\ee
where we have suppressed the arguments $(x,t)$ of the fields on the
right hand side of the equation.  The density field $\rho_{_{XP}}(\xi,t)$
satisfies a simple diffusion equation, so that its inhomogeneities
give rise to a diffusion current $j$.  We write
\be
j = - D {\partial \rho_{_{XP}} \over \partial\xi} = -D \left[{1\over 2} +
\rho_1 + \rho_2 - {\rho_1 \over 2} \, \ell'_1 (m_1) \right]^{-1}
\,\, {\partial \rho_{_{XP}} \over \partial x}
\ee
Here the diffusion constant $D$ is independent of densities $\rho_1$
and $\rho_2$.  Finally, in the current $j$, the odd and even zeroes
move with some local drift velocity, but their densities are in the
ratio $\rho_1:\rho_2$.  So the equation of motion for the fields $m_1$
is 
\be
{\partial m_1 \over \partial t} = - j_1(x,t) = {-\rho_1 \over \rho_1 +
\rho_2} \, j(x,t)
\ee
where $j(x,t)$ is given by Eq. (9.8), and a similar equation holds for
$m_2$.  Differentiating (9.4) with respect to $x$, we get
\be
\ell'_1(m_1) \cdot \rho_1(x,t) = \ell'_2(m_2) \cdot \rho_2 (x,t) 
\ee
Using this equation, it is easy to verify that the evolution equation
for $m_1$ and $m_2$ maintain the constraint condition Eq. (9.4), as
they should. 

We see that these equations differ from the usual hydrodynamical
equations in that the arbitrary functions $\ell_1(m)$ and $\ell_2(m)$
appear explicitly in the equations of motion.  These functions specify
the IS, and are completely determined by the initial conditions.  The
existence of the infinity of constants of motion given by the IS manifests
itself in hydrodynamics as these arbitrary functions which are
specified by the values of the constants of motion.  For the density
fields $\rho_1$ and $\rho_2$, the hydrodynamical description is seen
to depend only on the `classical' conservation law of IS.  The
observable consequences, if any, of the quantum-mechanical 
constants of motion of our model for the hydrodynamical description are
not very clear at present. 


\section{Summary and Concluding Remarks}

In summary, we have introduced a model of diffusing,
reconstituting dimers on a line.  The dynamics of 
this model satisfies the MSD property i.e. it is strongly non-ergodic, 
and the phase space can be decomposed into an exponentially large
number of mutually disconnected sectors. We determined the sizes and
numbers of these exactly.  We showed that these sectors could be distinguished
from each other by different values of a conserved quantity, the
Irreducible String.  The exact equivalence of the model to a model of
diffusing hard-core random walkers with conserved spin allowed us to
determine the sector-dependent behaviour of time-dependent correlation
functions in different sectors.  In any given sector, we showed that
the stochastic rate matrix was equivalent to the quantum Hamiltonian 
of a spin$-1/2$
Heisenberg chain (whose length depends on the sector), and thus demonstrated
that it was exactly diagonalizable. 

The equivalent quantum Hamiltonian $\hat H$ for the DRD model is related 
to an integrable quantum spin chain studied earlier by Bariev.  
The Bariev model has the Hamiltonian given by
\be
\hat H_B = - {1\over 4} \sum_i(\sigma^x_{i-1} \sigma^x_{i+1} +
\sigma^y_{i-1} \sigma^y_{i+1}) (1 - U\sigma^z_i)
\ee
The sign of $U$ can be changed by the transformation $\sigma^x_i
\rightarrow \sigma^y_i$, $\sigma^y_i \rightarrow \sigma^x_i,$ and 
$\sigma^z_i \rightarrow
-\sigma^z_i$.  We set $|U| = 1$.  Then the Bariev model differs from
our model through the term
\be
\hat H - \hat H_B = H' = - {1\over 4} \sum_i 
\left(\sigma^z_{i-1}\sigma^z_{i+1} -
1\right) \left(1+\sigma^z_i\right)
\ee
It is easy to see that these terms commute with the IS 
operator $\hat I_L(\lambda)$.  Thus $\hat I_L(\lambda)$ also commutes
with $\hat H_B$, and provides an infinity of constants of motion of the
Bariev model (in the special case $|U| = 1$).

The dynamics of the DRD model can also be viewed as a matrix
generalization of a one-dimensional KPZ-like surface roughening model,
in which the scalar height variable at site $i$ is replaced by matrix
valued variables $I_i(\lambda)$.  This matrix generalization differs
from that studied in \cite{moore,kardarzee}.  The (matrix) valued
height variable $I_i$ at point $i$ is defined in terms of the $A(1)\/$
and $A(0)\/$ matrices defined in Eq. (6.4) by
\be
I_i(\lambda) = \prod^i_{j=1} A(n_j,\lambda)
\ee
where the matrix product is ordered from left to right in order of
increasing $j\/$.  Using 
$A^2(1,\lambda) = 1$, it is easy to see that the matrix $I_i(\lambda)$
specifies the IS corresponding to the configuration to the left of and
including site $i$.  The stochastic evolution of the model is local in
the variables $\{I_i\}$:  If at any time $t$, $I_{i-1} = I_{i+1}$, then
with rate 1 we change either $I_{i-1}$ and $I_i$ to new values
$I'_{i-1}$ and $I'_i$, where
\be
I'_{i-1} = I_{i-2} A(1), \,\, I'_i = I_{i-2}
\ee
or change $I_i$ and $I_{i+1}$ to
\be
I''_i = I_{i+2},\, I''_{i+1} = I_{i+2} A(1),
\ee
leaving other matrices unchanged.  The height at the end site $i=L$ is
never changed, as it is just the constant of motion discussed earlier
[Eq. (6.3)].

The construction of the irreducible string in this model is
similar to other one-dimensional stochastic models with the
many-sector-decomposability property studied recently where this
construction has been found useful, i.e. the $k$-mer
deposition-evaporation model \cite{bar1,bardha} and the $q$-color dimer
deposition-evaporation model \cite{haridhar,thesis}
. (In the $q$-colour DDE model sites can be occupied by particles
of $q$ different colours. The update move consists of
changing the state of a pair of adjacent occupied sites of the same 
colour jointly to a
different colour.)  In all of these models, the long-time behaviour of the
time-dependent correlation functions can be obtained by assuming that it is
qualitatively the same as that of the spin-spin autocorrelation function in the
HCRWCS model with the same spin sequence as the IS in
the corresponding sector. 

However, the DRD model differs from earlier studied models in
significant ways.  In the trimer deposition-evaporation (TDE) model,
the correspondence between the configurations on the line and the position
of hard core walkers is many to one, unlike the present model where it
is one-to-one.  As a consequence, in the steady state of the TDE model,
all configurations of the random walkers are not equally likely, and
one has to introduce an effective interaction potential between the
walkers which is found to be of the form
\be
V(\{X_i\}) = \sum_i f(X_{i+1} - X_i)
\ee
where $\{X_i\}$ are the positions of the walkers, and $f(x)$ increases
as 
${3 \over 2} \ln X$ for large $X$\cite{bardha}.

In the TDE model, the transition probabilities for the random
walkers are also not completely independent of the spin-sequence of the
walkers.  In the $q$-color dimer deposition evaporation (qDDE) model,
the color symmetry of the model implies that the dynamics of random
walkers is completely independent of the spin sequence of the walkers,
but the potential of interaction $V$ is still present, which makes the
problem difficult to study exactly.  The present model is thus simpler
than both the TDE and the qDDE models, and has the additional virtue
of being exactly solvable in the sense that the stochastic matrix can
be diagonalized completely.

There are some straightforward but interesting generalizations of
the model.  Consider a general exclusion process with $k$ types of
particles.  In this general model, if we assume that some types of
particles cannot exchange positions (setting their exchange rate to
zero), their relative order will be conserved and this
can be coded in terms of the conservation of an IS.
As a simple example, consider a model with 4 species of particles
labelled A,B, C and D respectively with the allowed exchanges with
equal rates
\be
\begin{array}{l}
AC \rightleftharpoons CA, \;\;\; AD \rightleftharpoons DA \\
BC \rightleftharpoons CB  \;\;\; BD \rightleftharpoons DB.
\end{array}
\ee
This model again has the MSD property.  It
is easy to see that there are now {\it two} irreducible strings
which are conserved by the dynamics.  This is because the dynamics
conserves the relative order of A and B type particles, and also
of C and D type particles.  In a string specifying the
configuration formed of letters A,B,C and D, deleting all occurrences
of A and B gives rise to an IS specifying the relative
order between C and D type particles, which is a constant of motion.
Similarly, deleting all occurrences of C and D characters, we get
another independent IS which is also a constant of
motion.  In a specific sector, where both irreducible strings are
known, the dynamics treats A and B particles as indistinguishable,
as also  C and D.  Thus the dynamics is the same as that of the simple
exclusion process (with only two species of particles), and is
equivalent to the exactly solved Heisenberg model. 

Another generalization of the model would be to make the diffusion
asymmetric.  The corresponding 3-species exclusion process then
becomes asymmetric, and belongs to the KPZ universality class\cite{KPZ}.  The
corresponding stochastic matrix is again reducible to a simple
asymmetric exclusion process of 2 species, known to be exactly soluble
by Bethe ansatz techniques \cite{gwaspohn,halpin,kim},
and has a non-classical dynamical exponent 3/2.  The correlation
function of the asymmetric DRD model would map to somewhat complicated 
multispin correlation functions of the simple asymmetric exclusion 
process.  How these would vary from sector to sector has not been 
studied so far.

The DRD process in higher dimensions is also of interest.  For
example, it is easy to see that on a square lattice in two dimensions,
the number of totally jammed configurations increases
exponentially with the number of sites in the system.  All
configurations with no two adjacent 1's are totally jammed.  These are
just the configurations of the hard-square lattice gas model\cite{baxter},
whose number is known to increase exponentially with the area of
the system.  One can also construct configurations in which almost all
sites are jammed, except for a small number of diffusing dimers,
which can move only
along a finite set of horizontal or vertical lines 
(Fig 5).  Clearly the number of such sectors also increases as 
exponential of the area of the system.  In unjammed sectors, 
however, the dynamics is in general quite
nontrivial, and
there is no equivalence to the 2-d Heisenberg model.

\begin{figure}[htb]
\centerline{
        \epsfxsize=12.0cm
        \epsfysize=10.0cm
        \epsfbox{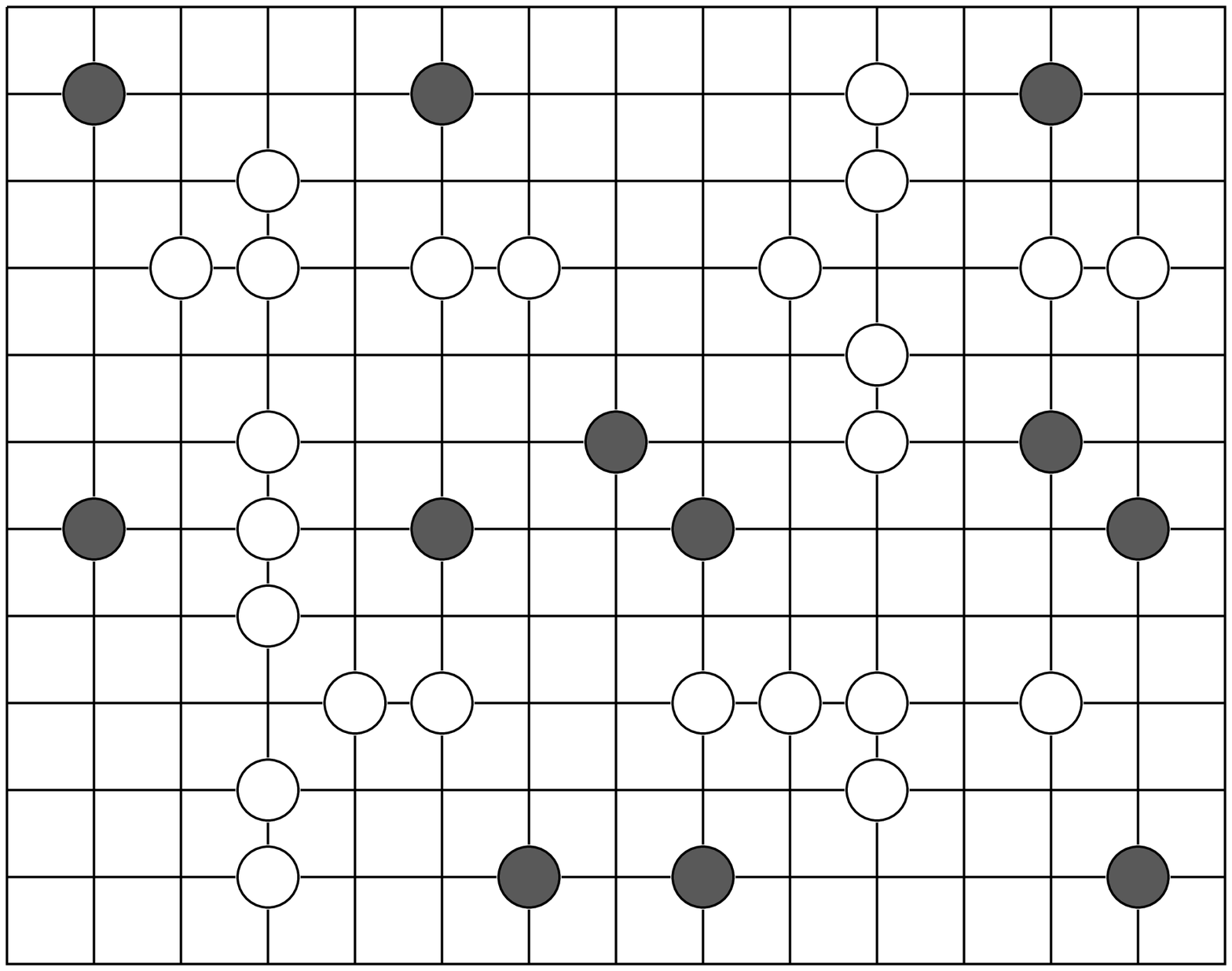}
}
\caption{An example of a partially jammed configuration of the
DRD model in two dimensions: atoms (denoted by white circles), on a
finite number of horizontal or vertical lines can move, while all
others (black circles) are completely immobile.}
\label{fig:5}
\end{figure}

An unexpected offshoot of our study was the construction of an
infinity of conservation laws for the special case of the $U=-1$ Bariev
model, from the construction of the irreducible string.  It would be
interesting to see if similar constructs can be found for other
quantum Hamiltonians, or for other stochastic evolution models having
the MSD property.

\end{document}